\documentclass[12pt,preprint]{aastex}
\usepackage{graphics,graphicx}
\def \cm{~\rm{cm}}
\def \s{~\rm{s}}
\def \km{~\rm{km}}

\def \G{~\rm{G}}
\def \AU{~\rm{AU}}
\def \erg{~\rm{erg}}

\def \yr{~\rm{yr}}
\def \pc{~\rm{pc}}

\def \lesssim{\mathrel{<\kern-1.0em\lower0.9ex\hbox{$\sim$}}}
\def \gtrsim{\mathrel{>\kern-1.0em\lower0.9ex\hbox{$\sim$}}}

\begin{document}

\title{Constraining the X-ray Luminosities of Asymptotic
  Giant Branch Stars: TX Cam and T Cas}

\author{Joel H. Kastner\altaffilmark{1} and Noam Soker\altaffilmark{2}}

\altaffiltext{1}{Chester F. Carlson Center for Imaging
Science, Rochester Institute of Technology, 54 Lomb Memorial
Dr., Rochester, NY 14623; jhk@cis.rit.edu}
\altaffiltext{2}{Department of Physics, Technion-Israel
Institute of Technology, Haifa 32000, Israel, and
Department of Physics, Oranim; soker@physics.technion.ac.il}

\begin{abstract}
To probe the magnetic activity levels of asymptotic giant
branch (AGB) stars, we used XMM-Newton 
to search for X-ray emission from two well-studied
objects, TX Cam and T Cas. The former star displays polarized maser
emission indicating magnetic field strengths of $B\sim5$ G;
the latter is one of the nearest known AGB stars. Neither
star was detected by XMM-Newton. We use the upper 
limits on EPIC (CCD detector) count rates to constrain
the X-ray luminosities of these stars, and derive
$L_X < 10^{31}$ erg s$^{-1}$ ($<10^{30}$ erg s$^{-1}$)
for an assumed X-ray emission temperature $T_X =
3\times10^6$ K ($10^7$ K). These limits represent $\lesssim 10$\%
($\lesssim 1$\%) of the
X-ray luminosity expected under models in which AGB magnetic
fields are global and potentially play an important role in collimating
and/or launching AGB winds. We suggest, instead, that 
the $B$ field strengths inferred from maser observations are
representative of localized, magnetic clouds.
\end{abstract}


\section{INTRODUCTION}

There has been considerable recent debate over the potential
role of magnetic fields in launching and/or shaping winds
from asymptotic giant branch (AGB) stars (Soker \& Kastner 2003 and
references therein). Of special interest are detections of maser
polarization around some AGB stars, which are 
indicative of the presence of relatively strong magnetic fields 
(e.g., Zijlstra et al.\ 1989; Kemball \& Diamond 1997;  
Miranda et al.\ 2001; Vlemmings, Diamond, \& van Langevelde \ 2002;
Bains 2004). SiO maser polarization measurements are
particularly important in this 
regard, as these observations probe the important transition region
between the AGB stellar photosphere and the wind. Based
on high-resolution polarization maps of SiO masers, 
Kemball \& Diamond (1997) deduce a magnetic 
field of $B \simeq 5-10 \G$ 
at a radius of $\sim 3.5 R_\ast$ around the AGB star TX Cam,
a Mira variable. Here $R_\ast$ 
is the stellar radius, which we take to be $\sim 2 \AU$
(see discussion in Kemball \& Diamond 1997). 
With a dependence on radius of $B \propto r^{-2}$ 
(Vlemmings et al. 2002) this suggests a stellar
surface magnetic field of $B_\ast \simeq 50-100 \G$. 
Miranda et al.\ (2001) find polarization in
the 1,665-MHz OH maser line that indicates the presence of
$\sim 10^{-3} \G$ magnetic fields at $\sim 10^{16} \cm$ from
the central star of the young planetary nebula (PN) K3-35.

Miranda et al.\ (2001) claim that their results favor
magnetic collimation models of outflows  
in PNs. The results for TX Cam, which will presumably
undergo a PN phase, could be similarly
interpreted to suggest that such magnetic collimation begins
well before ionization of the circumstellar
envelope. Indeed, in summarizing water maser polarization
measurements for several giants, Vlemmings et al.\ (2002) 
conclude that magnetic fields are strong enough to drive
and shape winds from AGB stars. 

However, these observations of
polarized maser emission also could
indicate the presence of localized, highly magnetized wind 
clumps, analogous to magnetic clouds in the solar wind 
(Soker \& Kastner 2003), rather than large-magnitude global
magnetic field strength (Soker 2002, 2003; Soker \& Kastner 2003; 
Soker \& Zoabi 2002).
In that respect we note the recent results of Murakawa et al.
(2003), who find that the H$_2$O maser clouds around the red
supergiant VX Sgr are $\sim 300$ times denser than the
surrounding wind. On the theoretical side, based on a dynamo model for the  
cool supergiant Betelgeuse, Dorch (2003)
finds that the magnetic structure has
a typical scale of $\sim 0.15 R_\star$,  smaller
than the the giant convection cells.

Soker \& Zoabi (2002) summarize possible problems in models where the
magnetic field plays a dynamical role in shaping the AGB wind.
Among others, they consider the X-ray luminosity ($L_X$).
As in the sun, they argue, globally strong magnetic fields will
violently reconnect, generating flares that lead to strong X-ray 
emission. Such a close coupling between 
stellar magnetic flux and X-ray luminosity has been
demonstrated to extend over 12-13 orders of magnitude in
$L_X$ (Pevtsov et al.\ 2003). 
Soker \& Zoabi find that for $B_\ast \gtrsim 1 \G$ on AGB stars, 
the expected X-ray luminosity is $\gtrsim 10^4$ times stronger 
than that of the Sun, if the reconnection rate per
unit surface area is similar to that in the Sun.
Likewise, if the X-ray luminosity is proportional to
the optical luminosity, the same scaling factor (from solar
to AGB X-ray luminosity) holds.

One might even expect that AGB star X-ray surface fluxes are
disproportionately larger than solar. This expectation is
based on the fact that for
the Sun --- where the mass loss rate is governed by
magnetic activity --- the average X-ray luminosity is of the same order
of magnitude as the rate of kinetic energy carried by the wind.
The X-ray luminosity in the soft X-ray band (that of the
ROSAT Position Sensitive 
Proportional Counter [PSPC], i.e., 0.1--2.4 keV) is in the range 
$\sim 3 \times 10^{26}$ to $\sim 5 \times 10^{27} \erg \s^{-1}$,
at minimum and maximum, respectively (Peres et al.\ 2000).
The solar wind's kinetic energy falls between these values.
If this is the case for AGB stars, as proposed in the dynamic-magnetic
models, then the X-ray luminosity should be a factor of
$\sim 10^6-10^8$ stronger than in the Sun (Soker \& Zoabi 2002).
Soker \& Zoabi conclude that this expectation, of $L_x \sim
10^{31} - 10^{35} \erg \s^{-1}$, is in sharp contradiction
with most ROSAT observations, which demonstrate that the 
X-ray luminosities of red giant stars typically 
are only marginally larger than the solar X-ray luminosity
(Schr\"oder {\it et al.} 1998) and that $L_x$ likely further decreases in
late giant evolution (H\"unsch \& Schr\"oder 1996). On the
other hand, some red giants are known to be relatively luminous
in X-rays (i.e., $L_x \sim 10^{30} \erg
\s^{-1}$; e.g., H\"unsch \& Schr\"oder 1996; H\"unsch 2001).

In Soker \& Kastner (2003) we discuss flares on AGB stars
from locally, rather than globally, strong fields. Such fields
should result in much weaker X-ray emission (Soker \& Zoabi
2002; Pevtsov et al.\ 2003). It is possible
that such weak coronal X-ray emission from heavily obscured,
mass-losing AGB stars has
escaped detection thus far, due to ROSAT's lack of hard X-ray
sensitivity and the large distances to these
short-lived, luminous stars. Indeed, Mira itself is a
weak ROSAT X-ray source ($L_x \sim 2\times10^{29} \erg
\s^{-1}$; Karovska et al.\ 1996; Soker \& Kastner 2003),
although the origin of its X-ray 
emission is uncertain given the presence of a close
companion. 

In an attempt to further constain models for
magnetic fields in AGB stars, we are conducting observations
of selected AGB stars with the XMM-Newton X-ray observatory,
which features sensitivity and energy coverage far superior
to that of ROSAT.
In this paper we report on and discuss results from
XMM-Newton observations of
the polarized maser source TX Cam and the nearby AGB star T Cas.
Like TX Cam, T Cas is a relatively strong SiO maser source (Herpin et al.\
1998), although we are unaware of measurements of the polarization of its
maser emission or of high-resolution imaging of its maser spots.

\section{Observations and Results}

We observed the TX Cam and T Cas fields with
XMM-Newton\footnote{For detailed
information concerning 
XMM-Newton and its instrumentation, see \\
http://xmm.vilspa.esa.es/external/xmm\_user\_support/documentation/index.shtml} 
on 2003 Sept.\ 4 and Feb.\ 6, respectively. The 
observatory's three coaligned telescopes provide imaging in
the $0.1-15$ keV energy range
with field of view of $\sim25'$ and 50\% encircled
energy diameter of $15''$. The instrument 
of interest for these observations was EPIC, 
with its three photon-counting CCD detector systems (pn, MOS 1, and MOS
2). The spectral resolution of these CCD systems range from
$\sim50$ eV to $\sim150$ eV over the energy range of
interest ($0.1-10$ keV). The thick filter was used for all
three detectors, to suppress detection of visible-light photons
from these optically bright targets.
Total EPIC integration times, broken down by detector, are
listed in Table 1.  

The 0.2-10 keV background count rate for each observation with a given
detector (pn, MOS 1, or MOS 2) was obtained 
from the total number of counts within an annulus centered on the
stellar position, where the annulus extends from $35''$ to
$70''$ in radius.  These measured background count
rates (Table 1) are consistent with those expected from the internal
``quiescent'' EPIC background combined with small
contributions from external background sources (see the
XMM-Newton Users' Handbook).  

The count rates within $18''$
radius circular source extraction regions centered on the
stellar positions were found to be consistent with the
surrounding background rates.  We conclude that no X-ray
sources were detected above background at the optical
positions of TX Cam and T Cas in these observations.  We
then obtain $3 \sigma$ upper limits on the 0.2-10 keV count
rate for each source from the count rate variances within
the source extraction regions, based on Poisson counting
statistics (Table 1).

While no X-ray source was detected at the position of either 
AGB star, these observations did yield the first-time detection
of $\sim12$ and $\sim40$ X-ray sources in the TX Cam and T
Cas fields, respectively. The identification of these
sources, most of which have no catalogued optical
or infrared counterparts, will be the subject of a future paper.

\begin{deluxetable}{cccc}
\tablewidth{0pt}
\tablecaption{\sc TX Cam and T Cas: XMM-Newton Observation Summary}
\tablehead{
\colhead{\sc Instrument}&
\colhead{\sc Int.\ Time }&
\colhead{\sc Bkdg.\ Rate\tablenotemark{a} }&
\colhead{\sc Src.\ Rate\tablenotemark{b} }
\\
\colhead{}  &
\colhead{(ks)}& 
\colhead{(ks$^{-1}$ arcsec$^{-2}$)}&
\colhead{(ks$^{-1}$)}
}
\startdata
\multicolumn{4}{c}{\sc TX Cam} \\
pn    & 6.929  & $2.7\times10^{-2}$ & $<7.5$ \\
MOS 1 & 13.468 & $3.9\times10^{-3}$ & $<3.5$ \\
MOS 2 & 13.488 & $3.0\times10^{-3}$ & $<3.6$ \\
\multicolumn{4}{c}{\sc T Cas} \\
pn    & 10.440 & $1.5\times10^{-2}$ & $<5.0$ \\
MOS 1 & 13.134 & $3.6\times10^{-3}$ & $<3.3$ \\
MOS 2 & 13.140 & $3.7\times10^{-3}$ & $<3.3$ \\
\enddata

\tablenotetext{a}{Background count rate (0.2-10 keV) obtained within
an annulus extending from $35''$ to $70''$
in radius, centered on stellar position. See text.}
\tablenotetext{b}{$3 \sigma$ upper limit (0.2-10 keV) on
  source count rate,
obtained by counting events within circular $18''$ radius
spatial region centered on stellar position. See text.}
\end{deluxetable}

\subsection{Upper Limits on $L_X$ for TX Cam and T Cas}

To derive upper limits on $L_X$ from the EPIC count rate
upper limits in Table 1, we used the
XSPEC\footnote{http://heasarc.gsfc.nasa.gov/docs/xanadu/xspec/}
software to compute intrinsic (unabsorbed) source fluxes for
a grid of representative $T_X$, $N_H$ values. We assumed a
standard Raymond-Smith plasma emission model (Raymond \&
Smith 1977) with
intervening absorption defined by the XSPEC ``wabs''
function (Morrison \& McCammon 1983), and used EPIC
spectral response matrices calculated for the specific
source extraction regions.

The models were constrained to reproduce the observed merged
MOS 1 and MOS 2 upper limit of $<2.5$ ks$^{-1}$
for TX Cam; the pn count rate upper limit is less useful,
due to the limited exposure time for the TX Cam observation.
Resulting upper limits on intrinsic source
X-ray flux $F_X$ are displayed in Fig.\ 1. To
calculate upper limits for $L_X$ from $F_X$, we take the
distance to TX Cam to be $320 \pc$ (Patel, Joseph, \&
Ganesan 1992). The limiting values for $F_X$ and
$L_X$ in Fig.\ 1 are
based on 3 $\sigma$ upper limits on count rate, and
therefore would become somewhat more stringent if we relax the
nondetection threshold. 

For T Cas, the EPIC upper limits are
marginally smaller, due to the additional pn integration
time, and Fig.\ 1 can be taken to represent somewhat more
conservative upper limits on $F_X$ from our XMM data for
this star. Note that T Cas may be only half as distant as TX
Cam (previous estimates range from 160 pc to 280 pc; Loup et
al.\ 1993), in which case the upper limits on $L_X$ implied
by our nondetection would be considerably more stringent than the
results for TX Cam in Fig.\ 1.

\section{DISCUSSION}

\subsection{Predicted X-ray Luminosity of TX Cam}

The X-ray luminosity due to magnetic activity associated
with an AGB star can be predicted in a variety
of ways, by analogy with the Sun. First, we can predict
$L_X$ from the kinetic energy carried by the
wind of the star, $\dot E_w$, assuming
the magnetic field is dynamically important (Soker \&
Zoabi 2002); in the case of the Sun, $L_X
\approx \dot E_w$. The wind speed of TX Cam is $v_w \sim 10
\km \s^{-1}$ and its mass loss rate
is $\sim10^{-6}$ (where the former quantity is obtained
directly from its millimeter-wave CO width and the latter is
obtained from a model of the CO line intensity and profile;
Knapp \& Morris 1985), suggesting $L_X \simeq \dot E_w 
\simeq 3 \times 10^{31} \erg \s^{-1}$.

For TX Cam, we can also predict $L_X$ from
the rate of magnetic energy carried by
the wind, $\dot E_B \simeq 4 \pi r^2 v_w(B^2/8 \pi)$, by
applying a scale factor between $\dot E_B$ and $L_X$ that is
obtained from the Sun. For the canonical solar surface magnetic
field value of $B \simeq 1 \G$ and wind speed
of $v \simeq 500 \km\s^{-1}$, we find $\dot E_B \simeq 10^{29} \erg
\s^{-1}$, which is $\sim 100$ times the X-ray luminosity of
the Sun. From the magnetic field strength inferred from maser
polarization measurements of TX Cam, $B
\simeq 5 \G$ at $r \simeq 7 \times 10^{13} 
\cm$ (Kemball \& Diamond 1997), we find $\dot E_B \simeq 6 \times
10^{34} \erg \s^{-1}$, if this field is carried
outward at $v \sim 10 \km \s^{-1}$ (this is an oversimplification, as the
kinematics of the maser spots is quite complicated; Diamond
\& Kemball 2001, 2003). Applying the solar scale factor between the rate of
magnetic energy loss and $L_X$ to TX Cam, we then predict $L_x
\sim 5 \times 10^{32} \erg \s^{-1}$, under the assumptions
that the TX Cam magnetic field is global and carried by the
AGB wind.

Both of these estimates, in turn, are similar
to what one would predict based on the relationship between
average (global) solar magnetic flux and solar $L_X$
(Pevtsov et al.\ 2003), and then 
scaling $L_X$ up according to the magnetic flux of TX Cam,
assuming a global magnetic field of 
$B \simeq 5 \G$ at $r \simeq 1000$ $R_\odot$.

We therefore estimate that the expected X-ray luminosity of
TX Cam is in the range $L_x \sim 3 \times 10^{31}- 5 \times
10^{32} \erg \s^{-1}$, for a globally and/or dynamically
important stellar magnetic field that is of the magnitude
measured from SiO maser polarization. Similar arguments
should pertain to T Cas, though estimates of $B$ are
unavailable for this star due to the lack of SiO polarization
measurements.

\subsection{X-ray Absorption by AGB Winds}

Given its mass loss rate of 
$\dot M \simeq 1.1 \times 10^{-6} M_\odot \yr^{-1}$
(Knapp \& Morris 1985), 
the wind of TX Cam could have a large column
density. Integrating along the line of site down to a radius
of $R_x$, and assuming the wind is expanding
with constant speed of
$v_w \sim 10 \km \s^{-1}$ and mass loss rate of
$\dot M \simeq 10^{-6} M_\odot \yr^{-1}$, the
total (ionized and neutral) H column density would be
\begin{equation}
N_H({\rm wind}) \simeq 10^{23}
\left( \frac {v_w}{10 \km \s^{-1}} \right)^{-1}
\left( \frac {\dot M}{10^{-6} M_\odot \yr^{-1}} \right)
\left( \frac {R_x}{2 \AU} \right)^{-1}      \cm^{-2} .
\end{equation}
Here $R_x$ is the radius where the X-ray emission by magnetic
activity takes place.
For such a large column density, the optical depth is
$\tau \sim 100$ (50) at $0.5$ keV ($1$ keV), assuming a wind
opacity similar to that of the ISM (see Draine \& Tan
2003). The mass-loss rate of T Cas is
$\sim3\times10^{-7} M_\odot \yr^{-1}$ and its wind speed is
$v_w \sim 6$ km 
s$^{-1}$ (Loup et al.\ 1993), yielding a similar
estimate for $N_H$ from Eq.\ 1 (assuming $R_x$ similar to
that of TX Cam).  

If Eq.\ 1 holds, we would not obtain meaningful upper limits
on $L_X$ from our nondetections of TX Cam and T Cas with
XMM-Newton (Fig.\ 1). There is reason to suspect, however,
that Eq.\ 1 substantially overestimates $N_H$. If $N_H \sim
10^{23} $ cm$^{-2}$, and the ISM relationship between $N_H$
and $A_V$ (e.g., Draine \& Tan 2003) applies here, Eq.\ 1
would suggest $A_V \sim 60$. Yet both stars are moderately
bright in the visual. For TX Cam (spectral type M8.5), $V$
varies between $\sim16.2$ and $\sim11.6$ (with a 557 day
period; Kukarkin et al.\ 1971). We then obtain a firm upper
limit of $A_V < 9$ by noting that $V-K < 16$ for TX Cam,
whereas $V-K > 7$ for very late-type M giants (Johnson
1966). Applying the same method to T Cas (M7e), which
displays $V$ of between $\sim12.4$ and $\sim7.3$ (Kukarkin
et al. 1971) and $V-K < 13$, we find $A_V < 6$. We conclude
that $N_H < 10^{22}$ cm$^{-2}$ for both stars, with $N_H$
toward T Cas somewhat smaller than toward TX Cam.

The discrepancy between this optically-derived
upper limit for $N_H$ and the estimate obtained via Eq.\ 1
likely reflects the fact that Eq.\ 1 relies on
the assumption of spherical symmetric
mass loss at constant rate. In fact, the combination of relatively bright
CO radio line emission and
moderate $A_V$ suggests either that (a) these stars are losing mass
primarily along their equatorial planes and are observed toward
relatively high latitudes along our line of
sight, or that (b) the relatively intense mass loss that resulted in their
expanding molecular envelopes occurred in short-lived episodes (as
has been observed in the case of many other AGB
stars). 

Furthermore, there is reason to expect that $N_H$ may be
somewhat smaller than either of the above estimates suggest.
If magnetic fields indeed shape the 
wind, we expect that some X-ray radiation will escape along
directions of lower density and/or lower opacity. Several
different models for non-spherical mass loss invoke magnetic
fields (e.g., Pascoli 1997; Matt et al.\ 2000;
Garc\'{\i}a-Segura \& L\'opez 2000; Blackman et al.\ 2001),
such that we might expect enhanced magnetic activity in
regions of lower density. In addition, the hot coronal gas formed by 
the magnetic activity above active regions could have much
reduced opacity  because many species will be highly
ionized, thus reducing the photoelectric absorption.
Finally, if the dust-to-gas ratios in the TX Cam and T Cas
envelopes are significantly larger than the ``canonical'' ISM value
of $\sim0.01$, the effective loss of metals from the gas phase likely would
further reduce $N_H$ (Wilms, Allen, \& McCray 2000).

\subsection{What is $T_X$ for TX Cam (and T Cas)?}

If $N_H \gtrsim 10^{22}$ cm$^{-2}$ and AGB magnetic reconnection events
yield circumstellar plasma conditions similar to those of
the solar corona ($T_X \sim2\times10^6$ K; e.g., Pevtsov et al.\
2003), then much of the resulting (soft) X-ray emission could be
attenuated by the AGB envelopes, and our nondetections would have less
significance (Fig.\ 1). However, while the correlation
between magnetic flux and $L_X$ is quite robust (Pevtsov et
al.\ 2003), the relationship (if any) between stellar
magnetic fields and the {\it temperature} of X-ray emission
has yet to be established. For stars of spectral type F and
later, it appears X-ray emission softens monotonically with
stellar age from pre-main sequence (T Tauri) through main
sequence stages (Kastner et al.\ 2003 and references
therein). But this X-ray spectral evolution may be due to
decreasing pre-main sequence disk accretion rates and/or
circumstellar $N_H$, or to changing plasma abundance
anomalies; and, in any event, there is no reason to expect
that the trend should continue into post-main sequence
evolution. Indeed, AGB star magnetic activity may have more
in common with T Tauri star activity than with main-sequence
activity, in which case we might expect relatively hard
X-ray emission and intense flaring (Feigelson et al.\
2002). In this regard it is intriguing that Mira displays
$T_X \sim 10^7$ K (Soker \& Kastner 2003), i.e., hotter than
the solar corona (although this emission may
originate from Mira's nearby companion; Karovska et al.\ 1996).

\section{CONCLUSIONS}

Given the above considerations, we tentatively place upper
limits of $L_X < 10^{31}$ erg s$^{-1}$ on the X-ray
luminosities of TX Cam and T Cas (Fig.\ 1). If the X-ray
emission due to magnetic activity is relatively hard, then
the upper limits become far more stringent: e.g., $L_X <
10^{30}$ erg s$^{-1}$ for $T_X = 10^7$ K and $N_H < 10^{22}$
cm$^{-2}$. 

In light of the large uncertainties, these results
do not rule out the  
claim that magnetic fields shape the winds from AGB stars.
However, it seems that the nondetection of TX Cam, in
particular, strongly
constrains such models. Given the evidence for relatively
strong magnetic fields around this star, and the tight
coupling between stellar magnetic flux 
and $L_X$ (e.g., Pevtsov et al.\ 2003), 
the upper limit on TX Cam's X-ray luminosity indicates that
the magnetic field  
behavior is much quieter and/or that its active regions are far
less widespread than on the Sun. Yet AGB stars, like TX Cam,
have very strong convective envelopes,  
strongly pulsate, and lose mass at high rates.
All these characteristics, we expect, should lead to a much more chaotic
magnetic field structure than on the Sun, with more
violent reconnection and liberation of magnetic energy
(Soker \& Kastner 2003). 

Our conclusion --- although not firm, due largely to
lingering uncertainties in $T_X$ and $N_H$ --- is that TX
Cam and T Cas are not strong X-ray sources, and hence their
average magnetic fields are much weaker than would be
required to collimate and/or drive their AGB winds. Instead,
the magnetic field must be concentrated in small regions,
some of which give rise to the polarized maser emission from
TX Cam (Soker \& Kastner 2003). Note that some magnetic
activity, but with much weaker X-ray  
luminosity than that expected for a global magnetic field, 
is still expected in our local field model (Soker \& Kastner 2003). 

Clearly more work, on both theoretical and observational
aspects of AGB star magnetic fields, is
needed. High-resolution polarization mapping of SiO emission
from T Cas and other relatively nearby, bright maser sources
is called for, as are additional, sensitive X-ray
observations of AGB stars with moderate $B$ fields, as inferred from
maser polarization and other methods.  We also suggest more
attention be paid to post-main sequence magnetic activity
models in which the fields are local, and have no global
dynamical role.

\acknowledgements{ The authors thank John Houck and the
anonymous referee for helpful suggestions concerning this paper.
We acknowledge support for this research
provided by NASA/GSFC XMM-Newton General Observer grant NAG5--13158 to RIT. 
N.S. acknowledges support from the Israel Science Foundation. }

\begin{figure}[htb]
\includegraphics[scale=1.,angle=0]{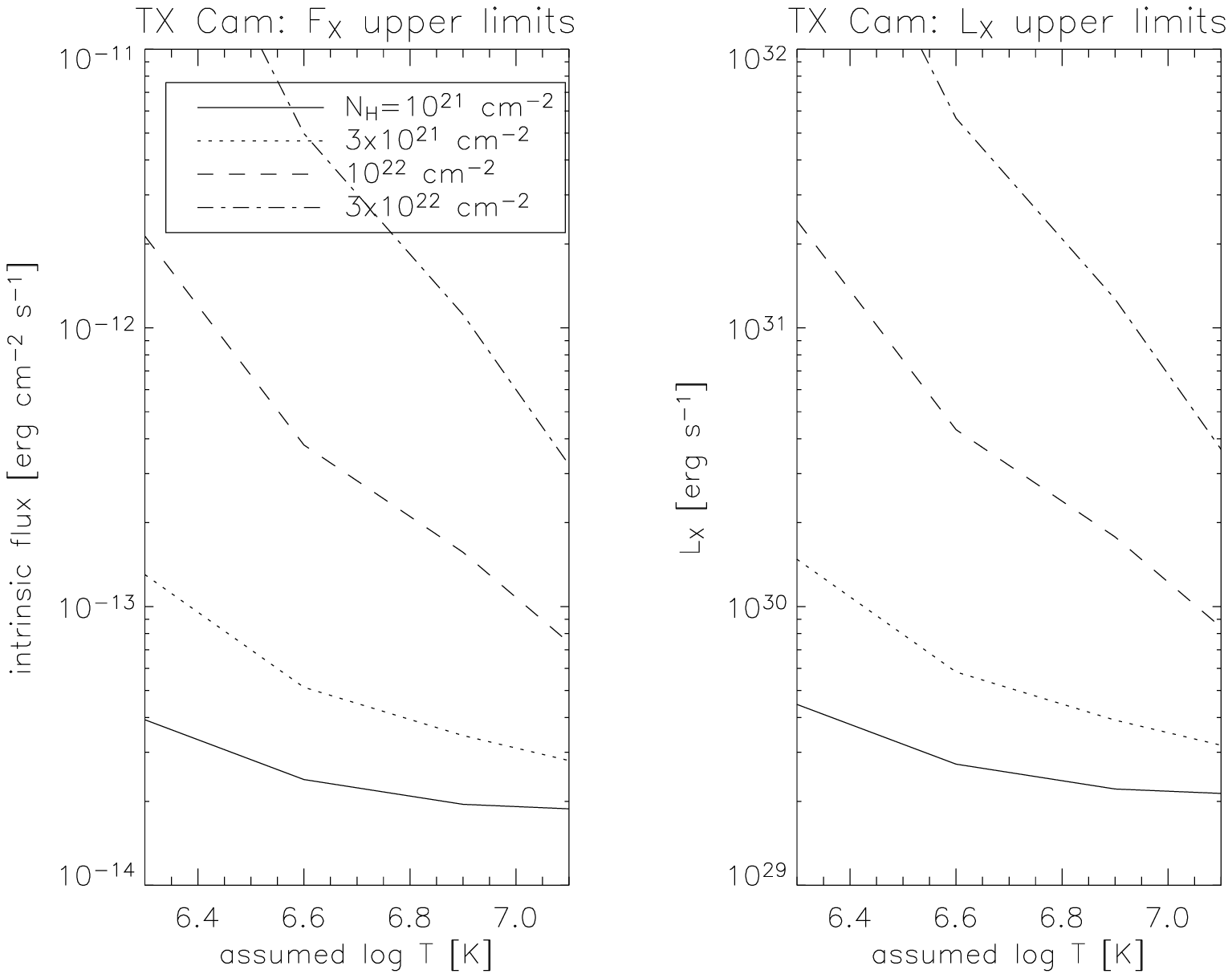} 
\caption{Upper limits on intrinsic X-ray fluxes (left) and source
  luminosities (right) imposed by the 3
  $\sigma$ upper limits on EPIC count rate observed for TX
  Cam, for Raymond-Smith coronal plasma models calculated
  over a range of assumed values of plasma temperature 
  ($T_X$) and intervening absorbing column ($N_H$).} 
\end{figure}

\end{document}